
\baselineskip=18pt
\documentstyle{amsppt}
\TagsOnRight



\nopagenumbers

\input amstex


\def\today{\ifcase\month\or January\or February\or March\or
April\or May\or June\or July\or August\or September\or
October\or November\or December\fi \space\number\day,
\number\year}


\def\gam{\gamma}

\def\kap{\kappa}
\def\lam{\lambda}


\def\calC{{\Cal C}}

\def\calM{{\Cal M}}




\font\tenboldgreek=cmmib10  \font\sevenboldgreek=cmmib10 at
7pt
\font\fiveboldgreek=cmmib10 at 7pt
\newfam\bgfam
\textfont\bgfam=\tenboldgreek \scriptfont\bgfam=\sevenboldgreek
\scriptscriptfont\bgfam=\fiveboldgreek


\font\tengerman=eufm10 \font\sevengerman=eufm7 \font\fivegerman=eufm5
\font\tendouble=msym10 \font\sevendouble=msym7 \font\fivedouble=msym5

\textfont4=\tengerman \scriptfont4=\sevengerman
\scriptscriptfont4=\fivegerman
\newfam\dbfam
\textfont\dbfam=\tendouble \scriptfont\dbfam=\sevendouble
\scriptscriptfont\dbfam=\fivedouble
\def\gr{\fam4}

\mathchardef\ng="702D
\mathchardef\dbA="7041
\mathchardef\sm="7072
\mathchardef\nvdash="7030
\mathchardef\nldash="7031
\mathchardef\lne="7008
\mathchardef\sneq="7024
\mathchardef\spneq="7025
\mathchardef\sne="7028
\mathchardef\spne="7029
\mathchardef\ltms="706E
\mathchardef\tmsl="706F

\mathchardef\dbA="7041

	
	\def\grb{{\gr b}}

	\def\grg{{\gr g}}
	\def\grh{{\gr h}}


\def\ZZ{\Bbb Z}

\def\NN{\Bbb N}


\def\sdp{\times \hskip -0.3em {\raise 0.3ex
\hbox{$\scriptscriptstyle |$}}} 


\def\End{\operatorname{End\,}}









\def\hatlam{{\widehat\lambda}}



\def\sqr#1#2{{\vcenter{\hrule height.#2pt\hbox{\vrule
width.#2pt height#1pt \kern#1pt \vrule width.#2pt}\hrule
height.#2pt}}}

\def\buildrul#1\under#2{\mathrel{\mathop{\null#2}\limits_{#1}}}

\def\boxit#1{\vbox{\hrule\hbox{\vrule\kern3pt\vbox{\kern3pt#1
\kern3pt}\kern3pt\vrule}\hrule}}

\def\subheading#1{\medskip\goodbreak\noindent{\bf
#1.}\quad}

\def\longmapright #1 #2 {\smash{\mathop{\hbox to
#1pt {\rightarrowfill}}\limits^{#2}}}
\def\longmapleft #1 #2 {\smash{\mathop{\hbox to
#1pt {\leftarrowfill}}\limits^{#2}}}

\def\back{{\raise 2.5pt\hbox{$\,\scriptscriptstyle\backslash\,$}}}

\def\part{\partial}
\def\lwr #1{\lower 5pt\hbox{$#1$}\hskip -3pt}
\def\rse #1{\hskip -3pt\raise 5pt\hbox{$#1$}}
\def\lwrs #1{\lower 4pt\hbox{$\scriptstyle #1$}\hskip -2pt}
\def\rses #1{\hskip -2pt\raise 3pt\hbox{$\scriptstyle #1$}}

\def\<#1{\left\langle{#1}\right\rangle}

\def\subinbn{{\subset\hskip-8pt\raise
0.95pt\hbox{$\scriptscriptstyle\subset$}}}

\def\llvdash{\mathop{\|\hskip-2pt \raise 3pt\hbox{\vrule
height 0.25pt width 1.5cm}}}

\def\lvdash{\mathop{|\hskip-2pt \raise 3pt\hbox{\vrule
height 0.25pt width 1.5cm}}}

\def\fakebold#1{\leavevmode\setbox0=\hbox{#1}%
  \kern-.025em\copy0 \kern-\wd0
  \kern .025em\copy0 \kern-\wd0
  \kern-.025em\raise.0333em\box0 }

\font\msxmten=msxm10
\font\msxmseven=msxm7
\font\msxmfive=msxm5
\newfam\myfam
\textfont\myfam=\msxmten
\scriptfont\myfam=\msxmseven
\scriptscriptfont\myfam=\msxmfive
\mathchardef\rhookupone="7016
\mathchardef\ldh="700D
\mathchardef\leg="7053
\mathchardef\ANG="705E
\mathchardef\lcu="7070
\mathchardef\rcu="7071
\mathchardef\leseq="7035
\mathchardef\qeeg="703D
\mathchardef\qeel="7036
\mathchardef\blackbox="7004
\mathchardef\bbx="7003
\mathchardef\simsucc="7025

\def\+{\oplus }
\def\x{\times }
\def\ten{\otimes}

\def\mqg{\calM _q(\grg)}

\def\uqg{U_q(\grg)}

\def\vl{V(\lam)}

\def\vnl{V(n\lam)}
\def\vmu{V(\mu)}
\def\vtn{V^{\otimes n}}

\topmatter

\title On quantum flag algebras
\endtitle

\author
Alexander Braverman
\endauthor

\address
School of Mathematical Sciences, Tel-Aviv University, Ramat-Aviv,
Israel
\endaddress

\email
braval\@math.tau.ac.il
\endemail

\abstract
Let $G$ be a semisimple simply connected algebraic group over an algebraically
closed field of characteristic $0$. Let $V$ be a simple finite-dimensional
$G$-module and let $y$ be its highest weight vector. It is a classical result
of B.~Kostant that the algebra of functions on the closure of $G\cdot y$ is
quadratic.
In this paper we generalize
this result to the case of the quantum group $\uqg$. The proof uses
information about $R$-matrix due to Drinfeld and Reshetikhin.
\endabstract
\endtopmatter

\heading 0. Introduction
\endheading
\subheading{0.1. Notations}Let $\grg$ be a semisimple Lie algebra over an
algebraically closed field $k$ of characteristic $0$. Let us choose a Borel
subalgebra $\grb\subset \grg$ and a Cartan subalgebra $\grh\subset\grb$.
Let $\Pi\subset \grh^*$ be the set of simple roots of $\grh$ with respect
to these choises. We shall denote by $P(\Pi)$ the corresponding weight lattice
and by $P(\Pi)^+$ the set of dominant weights in $P(\Pi)$. We shall denote by
$\rho\in P(\Pi)^+$ the half sum of all positive roots. For any
$\lam,\mu\in\grh ^*$ the symbol $\lam>\mu$ will mean that $\lam-\mu$ is
a sum of positive roots.  Choose an invariant bilinear form on $\grg$. We shall
denote by $(\cdot,\cdot)$ its restriction to $\grh$ (and to $\grh^*$ by
transport of structure) and by $\|\cdot\|$ the corresponding
norm on $\grh^*$.
\subheading{0.2}Let $K=k(q)$ where $q$ is transcendental over $k$. Let
$\uqg$ denote the quantized universal enveloping algebra of $\grg$
(cf. for example [D1], [L]) constructed using the invariant form of 0.1
(as in [D1]). For any
$\lam\in P(\Pi)^+$ we shall denote by $\vl$ the simple $\uqg$-module with
highest weight $\lam$ (cf. [L]). We shall also denote by $\mqg$ the tensor
category of locally finite integrable (cf. [L]) $\uqg$-modules.
\subheading{0.3. Quadratic algebras}For any $K$-vector space $V$ we shall
denote by $T(V)$ the tensor algebra of $V$. Let $A=\oplus _{n=0}^{\infty}A_n$
be a graded $K$-algebra.
\proclaim{Definition}We say that $A$ is {\it quadratic} if
\roster
\item $A_0=K$ and $A$ is generated by $A_1$
\item the ideal of relations of $A$ is quadratic, i.e. if we let $i$ denote the
natural map $T(A_1)\to A$ then
\endroster
$$
\ker i=\sum_{i,j\in\ZZ ^+}V^{\ten i}\ten W\ten V^{\ten j}
$$
where $W=\ker i\bigcap V^{\ten 2}$.
\endproclaim

\subheading{0.4. The result} Let $\lam\in P(\Pi)^+$
and let $V=\vl$. Let us define a graded algebra $A(V)=\oplus_{n=0}^{\infty}
A_n(V)$
by
$$
A_n(V):=V^{\ten n}/\sum_{\vmu\subset\vtn, \mu <n\lam}\vmu
$$
and multiplication coming from the tensor algebra of $V$.
In the classical ($q$=1) case the algebra $A(V)$ is isomorphic to the algebra
of functions on the closure of $G\cdot y$ where $G$ is the simply connected
algebraic group which corresponds to $\grg$ and $y$
is a highest weight vector in the $\grg$-module
$V^*$, dual to $V$ and hence can be identified with the coordinate ring
of the flag variety of $\grg$ with respect to the invertible sheaf defined
by $V$. This is why in our situation the algebra $A(V)$ deserves the name
of a quantum flag algebra.
\proclaim{Theorem}The algebra $A(V)$ is quadratic.
\endproclaim
In the classical case this the analogous theorem was proven by B.~Kostant
(cf. for example [FH]).

\subheading{0.5. Remarks}1) In fact one can easily see that the algebra
$A(V)$ is defined as an algebra in the category $\mqg$ and theorem 0.4
is equivalent to the statement that $A(V)$ is quadratic as an $\mqg$-algebra
(cf [HS] for the relevant definitions).

2) In the case $\grg=sl_n$ theorem 0.4 can be deduced from [TT].

\heading{1. Proof of the theorem}
\endheading
\subheading{1.1. The braiding}Let $(\calC, \ten)$ be a monoidal category (cf.
[SS]). A {\it braiding} on $\calC$ is an isomorphism $B$ of two functors
$\calC\x\calC\to \calC$ (namely, $(V,W)\to V\ten W$ and $(V,W)\to W\ten V$ for
any $(V,W)\in Ob~\calC\x\calC$), which satisfies the hexagon identities (cf.
[SS]). Drinfeld in [D1] has shown that $\mqg$ has a braiding and it was shown
by
D.~Gaitsgory ([G]) that this braiding was essentially unique.
For any $V,W\in Ob~\mqg$ we shall denote by $B_{V,W}$ the corresponding
morphism $B_{V,W}:\, V\ten W\to W\ten V$. Let also
$S_{V,W}=B_{W,V}\circ B_{V,W}\in \End V\ten W$.
\subheading{1.2}For any $\lam\in\grh ^*$ define $c(\lam)=(\lam, \lam +2\rho)=
\|\lam + \rho\|^2-\|\rho\|^2$. Let $\lam,\mu\in P(\Pi)^+$. For any
$\gam\in P(\Pi)^+$ denote by $P_{\lam ,\mu}^{\gam}\in \End (V(\lam)\ten
V(\mu))$
the projector from $\vl\ten \vmu$ to its $V(\gam)$-isotypical component.
The following statement is due to N.~Reshetikhin ([R], cf. also [D2, $\S$ 5]).
\proclaim{Lemma}
$$
S_{\vl,\vmu}=\sum _{\gam\in
P(\Pi)^+}q^{2(c(\gam)-c(\lam)-c(\mu))}P_{\lam,\mu}^{\gam}
$$
\endproclaim
\subheading{1.3}Let $V=V(\lam)$ and $S=S_{V,V^{\ten n-1}}$ (for some $n\in
\NN$).
\proclaim{Corollary}
$$
\{ x\in \vtn| Sx=q^{2(c(n\lam)-c(\lam)-c((n-1)\lam))}x\} =\vnl\subset \vtn
$$
i.e. $\vnl$ is a full eigenspace of $S$ with eigenvalue
$q^{2(c(n\lam)-c(\lam)-c((n-1)\lam))}$.
\endproclaim
\demo{Proof}Let $P_{n-1}=\{ \mu\in P(\Pi)^+ |\vmu\subset \vtn\} $. We know from
1.2 that $\vnl$ is an eigenspace of $S$ with eigenvalue
$q^{2(c(n\lam)-c(\lam)-c((n-1)\lam))}$.
On the other hand 1.2 again implies that any other eigenvalue of $S$ takes the
form $q^{2(c(\gam)-c(\lam)-c(\mu))}$ where $\mu\in P_{n-1}$ and
$V(\gam)\in V\ten \vmu$. Let us prove that if $\gam\neq n\lam$ or
$\mu\neq (n-1)\lam$ then
$$
c(\gam)-c(\mu)<c(n\lam)-c((n-1)\lam) \tag{$*$}
$$
(this clearly implies the corollary). The left hand side of ($*$) is equal to
$\|\gam+\rho\|^2-\|\mu+\rho\|^2$. Clearly $\gam\leq\mu +\lam$, so, it is enough
to prove ($*$) only in the case when $\gam=\mu +\lam$. In this case
$$\align
[c(n\lam)-c((n-1)\lam)]&-[c(\gam)-c(\mu)]=\\
(\|n\lam +\rho\|^2-\|(n-1)\lam +\rho\|^2)&-(\|\mu+\lam
+\rho\|^2-\|\mu+\rho\|^2)=\\
2((n-1)\lam+\rho,\lam)-2(\mu+\rho,\lam&)=2((n-1)\lam-\mu,\lam)
\endalign
$$
Let $\nu =(n-1)\lam-\mu$.
Since $\lam$ is dominant and $\mu \leq (n-1)\lam$, we have
$(\nu,\lam)\geq 0$. Suppose that $(\nu,\lam)=0$. Then
$$
0\leq (\nu,\mu)=(\nu,(n-1)\lam-\nu)=(n-1)(\nu,\lam)-\|\nu\|^2=
-\|\nu\|^2<0
$$
and therefore $\nu=0$ and $\gam =n\lam $ and $\mu =(n-1)\lam$ which is a
contradiction to our assumption. $\square$
\enddemo

\proclaim{1.4. Lemma}$B_{V,V}|_{V(2\lam)\subset V\ten V}=\kap \circ
id_{V(2\lam)}$ for some $\kap\in K$.
\endproclaim
\demo{Proof}This follows from the fact that $B_{V,V}$ is a morphism of modules
and that $V(2\lam)$ has multiplicity one in $V\ten V$.
\enddemo
\subheading{1.5}Now we are ready to prove theorem 0.4. Let $\hatlam$ denote the
highest weight of the module $V^*$, dual to $V$. We are going to prove
the quadraticity of the algebra $A(V^*)$. Let
$W=\sum_{\vmu\subset V^*\ten V^*,\mu<2\hatlam}\vmu\subset V^*\ten V^*$. Then
the quadraticity of the
algebra $A(V^*)$ is equivalent to
$$
\sum_{i=1}^{n-1}V^{*\ten i-1}\ten W\ten V^{*\ten n-i-1}=
\sum_{\vmu\subset V^{*\ten n},\mu<n\hatlam}\vmu\subset V^{*\ten n}\tag{$**$}
$$
Let $V^n_i=V^{\ten i-1}\ten V(2\lam)\ten V^{\ten n-i-1}\subset \vtn$. Then
the orthogonal complement to the left hand side of ($**$) in $\vtn$ is equal to
$\bigcap _{i=1}^n V^n_i$ and the orthogonal complement to the right hand side
of ($**$) is equal to $\vnl$. Hence ($**$) is equivalent to the following
\proclaim{Proposition}$\bigcap _{i=1}^n V^n_i=\vnl$
\endproclaim
\demo{Proof}After corollary 1.3 it is enough to check that the left hand side
is
an eigenspace of $S$ (since we have obvious embedding $\vnl\subset\bigcap
_{i=1}^n V^n_i$).
For any $1\leq i\leq n-1$ denote by $B^{i,i+1}$ the morphism
$id_{V^{\ten i-1}}\ten B_{V,V}\ten id_{V^{\ten n-i-1}}$. Then the hexagon
identities imply that $B_{V,V^{\ten n-1}}=B^{n-1,n}\circ ...\circ B^{1,2}$
and $B_{V^{\ten n-1},V}=B^{1,2}\circ ...\circ B^{n-1,n}$. Let $\kap$ be as
in 1.4. Then
$$
\align
S|_{\bigcap _{i=1}^n V^n_i}=
B_{V^{\ten  n-1},V}\circ B_{V,V^{\ten n-1}}|_{\bigcap _{i=1}^n V^n_i}=&\\
B^{1,2}\circ B^{2,3}\circ ...\circ B^{n-1,n}\circ B^{n-1,n}\circ ...
\circ B^{2,3}\circ B^{1,2}|_{\bigcap _{i=1}^n V^n_i}&=
\kap ^{2(n-1)}\circ id _{\bigcap _{i=1}^n V^n_i}
\endalign
$$
which finishes the proof. $\square$
\enddemo

\subheading{1.6. Concluding remarks}In [Be] R.~Bezrukavnikov extended Kostant's
result, showing that for $q=1$ not only is the algebra $A(V)$ quadratic, but it
is also a  Koszul algebra (cf. for example [BG]). It would be interesting to
extend this result to the quantum case. One idea in this direction could be the
following. It is remarked in [BG] that if $V$ is a vector space and
$B\in \End V$ is a Yang-Baxter operator (cf. [D1] or [BG] ) which is
unitary (i.e. $B^2=id$) then the subspace
$\{ x\in V\ten V|\ Bx=x\} $ of $V\ten V$ defines a Koszul algebra. In our
situation one can show that in fact $V(2\lam)\in V\ten V$ is an eigenspace
of $B_{V,V}$. However $B_{V,V}^2\neq id$, but one may try to prove the
statement
of [BG] holds for a broader class of Yang-Baxter operators, which will include
$B_{V,V}$. This will prove the Koszulity of the dual algebra of $A(V)$
and hence the Koszulity of $A(V)$ itself (we refer to [BG] for the definition
of duality for quadratic algebras and its relation to Koszulity).
\subheading{1.7. Acknowledgements}I am grateful to A.~Joseph for posing tis
problem and stimulating discussions. I also would like to thank J.~Bernstein,
D.~Gaitsgory and S.~Shnider for valuable conversations concerning the subject.

\enddocument